\documentclass{article}

\usepackage{microtype}
\usepackage{graphicx}
\usepackage{subfigure}
\usepackage{booktabs} 
\usepackage{hyperref}

\usepackage[accepted]{icml2024}

\usepackage{amsmath}
\usepackage{amssymb}
\usepackage{mathtools}
\usepackage{amsthm}

\usepackage[capitalize,noabbrev]{cleveref}

\theoremstyle{plain}

\theoremstyle{definition}

\theoremstyle{remark}

\icmltitlerunning{Designing an Evaluation Framework for LLMs in Astronomy Research}

\begin{document}

\twocolumn[
\icmltitle{Designing an Evaluation Framework for\\Large Language Models in Astronomy Research}

\icmlsetsymbol{equal}{*}

\begin{icmlauthorlist}
\icmlauthor{John F. Wu}{stsci,jhu}
\icmlauthor{Alina Hyk}{oregonstate} %
\icmlauthor{Kiera McCormick}{loyola}
\icmlauthor{Christine Ye}{stanford} %
\icmlauthor{Simone Astarita}{esa} %
\icmlauthor{Elina Baral}{jhu}
\icmlauthor{Jo Ciuca}{anu}
\icmlauthor{Jesse Cranney}{anu} %
\icmlauthor{Anjalie Field}{jhu} %
\icmlauthor{Kartheik Iyer}{columbia} %
\icmlauthor{Philipp Koehn}{jhu} %
\icmlauthor{Jenn Kotler}{stsci} %
\icmlauthor{Sandor Kruk}{esa} %
\icmlauthor{Michelle Ntampaka}{stsci} %
\icmlauthor{Charles O'Neill}{anu} %
\icmlauthor{Joshua E.G. Peek}{stsci} %
\icmlauthor{Sanjib Sharma}{stsci} %
\icmlauthor{Mikaeel Yunus}{jhu} %
\end{icmlauthorlist}

\icmlaffiliation{stsci}{Space Telescope Science Institute, Baltimore, MD, USA}
\icmlaffiliation{jhu}{Johns Hopkins University, Baltimore, MD, USA}
\icmlaffiliation{oregonstate}{Oregon State University, Corvallis, OR, USA}
\icmlaffiliation{loyola}{Loyola University Maryland, Baltimore, MD, USA}
\icmlaffiliation{stanford}{Stanford University, Stanford, CA, USA}
\icmlaffiliation{esa}{European Space Astronomy Centre, Madrid, Spain}
\icmlaffiliation{anu}{Australian National University, Canberra, Australia}
\icmlaffiliation{columbia}{Columbia University, New York, NY, USA}

\icmlcorrespondingauthor{John F. Wu}{jowu@stsci.edu}

\vskip 0.3in
]

\printAffiliationsAndNotice{}  %

\begin{abstract}
Large Language Models (LLMs) are shifting how scientific research is done. It is imperative to understand how researchers interact with these models and how scientific sub-communities like astronomy might benefit from them. However, there is currently no standard for evaluating the use of LLMs in astronomy. Therefore, we present the experimental design for an evaluation study on how astronomy researchers interact with LLMs. We deploy a Slack chatbot that can answer queries from users via Retrieval-Augmented Generation (RAG); these responses are grounded in astronomy papers from arXiv. We record and anonymize user questions and chatbot answers, user upvotes and downvotes to LLM responses, user feedback to the LLM, and retrieved documents and similarity scores with the query. Our data collection method will enable future dynamic evaluations of LLM tools for astronomy. %
\end{abstract}

\section{Introduction}

Scientific research traditionally involves reviewing the literature, obtaining and analyzing data, formulating hypotheses, and publishing results. Internet search engines, data archives, and other technological advancements have helped streamline and democratize the research process. Bibliographic systems like arXiv and the NASA Astrophysics Data System \citep[ADS;][]{ads} are critical for finding relevant publications and facilitating research in astronomy. 

The way we do research continues to evolve, especially since the advent of Large Language Models (LLMs). Researchers will still need to perform a literature review before embarking on research. Nevertheless, LLMs have the potential to make this process more powerful and efficient. For example, they can provide a natural language frontend to systems like arXiv in order to provide semantic search. LLMs can also retrieve papers and synthesize responses using the added information as context \citep[known as Retrieval-Augmented Generation, or RAG;][]{RAG}. 
RAG has been touted as a solution for mitigating ``hallucinations'' \citep{Ji_2023} and providing access to specialized, up-to-date, domain knowledge. 

But how do we know if these tools are 
actually helping scientific research? LLMs are notoriously brittle and can fall prey to surprising failure modes. As one example, RAG can be adversely affected by irrelevant or contradictory information in retrieved documents \citep{Gao2023RetrievalAugmentedGF}. %

Thus, we aim to understand how LLMs are used in scientific research. We see a need for dynamic evaluation studies, which can capture real-world interactions between users and LLMs, rather than static benchmarks. Research astronomy is particularly well-suited for conducting user evaluation studies due to its low risks, absence of personally identifiable information (PII), and open non-commercial data.

Our contributions are two-fold.
First, we engineer a LLM chatbot that generates responses to astronomy research questions by retrieving arXiv papers listed in the astro-ph (astrophysics) category. Users can interact with the RAG-based chatbot through a Slack application.
Second, we design a framework for evaluating the aforementioned LLM chatbot in an astronomy research setting. We can record user interactions with the LLM, providing a rich data set of (\textit{a}) user research questions and LLM answers, (\textit{b}) user upvotes and downvotes for the LLM answers, (\textit{c}) user feedback to the LLM, and (\textit{d}) retrieved documents and similarity scores.

Here, we present the experimental design for collecting (anonymized) user data in an upcoming study; we have not collected any data yet. We design this evaluation framework for a user base of  professional astronomy researchers (i.e., professional scientists with PhDs in physics/astronomy). In future works, we will present the compiled data, as well as the evaluation results. 
Our proposed evaluation study has been approved by an Institutional Review Board (IRB).

\section{Related Work}

Specialized LLMs fine-tuned for astronomy have recently emerged.  
\citet{Nguyen2023AstroLLaMA} release AstroLLaMA, a LLaMA-2 7B model fine-tuned using over 300,000 astronomy abstracts from arXiv; the authors find that AstroLLaMA surpasses GPT-3---a far larger foundation model---on text completion and embedding tasks for astronomy topics.
\citet{Perkowski2024} report that, while general LLMs such as GPT-4 excel in broader question-answering scenarios due to superior reasoning capabilities, continual pre-training of smaller astronomy-focused models, such as AstroLLaMA-chat, can enable competitive performance on specialized astronomy topics.

LLMs for astronomy also benefit from document retrieval.
\citet{Ciuca2023Galactic} showcase the effectiveness of in-context learning and RAG from astronomy papers to perform summarization, comparative analysis, and idea generation. %
\citet{Ciuca2023Harnessing} combine in-context learning with adversarial prompting for hypothesis generation. By giving GPT-4 access to papers in one specific astronomy subfield and allowing another model to act as an adversary, the authors show that model-generated scientific hypotheses improve in quality, as evaluated by human experts. %

LLMs are also useful for extracting structured information from astronomy papers or other text sources.
\citet{grezes2021building} present one of the earliest LLM applications in astronomy: astroBERT, a tool designed to perform named entity recognition for NASA ADS.
\citet{shao2024astronomical} evaluate several open-source and closed-source LLMs for astronomical named entity recognition/extraction.
\citet{Sotnikov2023Language} extract information from transient event notifications (ATel\footnote{\url{https://astronomerstelegram.org/}} and GCN\footnote{\url{https://gcn.nasa.gov/}} messages) by training a model on a small number of examples (few-shot learning) and through prompt engineering.
\citet{2024zndo..10738296V} release software that can parse astronomers' publications in order to identify their area of expertise.

\section{Generating robust answers with LLMs}

\subsection{Steering LLMs with prompting}

LLMs can be given some context alongside or prior to any user queries, a technique called prompting. One particularly useful prompt for scientific applications is to allow the LLM to say ``I don't know'' if the answer is uncertain. Other kinds of prompts can request that the LLMs explicate their chain of thought \citep{wei2023chainofthought}, which seems to be particularly effective for LLMs due to their autoregressive nature.

In-context learning is a related method for supplying demonstrations of good (or bad) responses to LLMs as part of the prompt. This approach seems to succeed because sufficiently large language models are zero-shot learners \citep{gpt3}.

\subsection{Information retrieval}
Information retrieval is an well-studied problem in language modeling and computer science, wherein a computer system is tasked with searching for information based on a user query. This is particularly relevant for scientific fields like astronomy, for which there exists a large and esoteric corpus of domain-specific knowledge that can be difficult to query. 

LLMs are often used as encoders for information retrieval due to their ability to represent the semantics of the user query and of other documents
\citep{reimers2019sentencebert,khattab2020colbert,bge_embedding}. Modern LLMs can be tasked with retrieving relevant passages, re-ranking initial retrieval results, summarizing documents, and synthesizing information from multiple sources \citep[e.g.,][]{DBLP:journals/corr/abs-1901-04085, Weller2023AccordingT,zhang2023benchmarking}.

\subsection{Retrieval Augmented Generation}
Through retrieval of external documents, LLMs can supplement their knowledge using  relevant information as additional context \citep{RAG}. RAG allows models to access information beyond the scope of their original training or fine-tuned datasets \citep[e.g.,][]{Taylor2022Galactica}, updating the model's knowledge base with new, private, or domain-specific information. 

RAG has been shown to reduce hallucinations and increase knowledge use in large language model outputs \citep{shuster2021retrieval,Ji_2023}. RAG-based systems have been successfully deployed in a number of domain-specific applications, such as clinical medicine and scientific research, and improve the robustness of generated responses \citep{lála2023paperqa, zakka2023almanac}.

\section{Experimental Design} \label{sec:design}

We present a LLM chatbot that retrieves information from arXiv astro-ph papers in order to answer user queries (Section~\ref{sec:rag-bot}). By default, this system uses \texttt{gpt-4o} as the generator LLM and \texttt{bge-small-en-v1.5} as the encoder LLM. We deploy this chatbot in Slack to facilitate user interactions and feedback, which is stored for future evaluation.

\subsection{RAG with astronomy arXiv papers} \label{sec:rag-bot}

\begin{figure*}[ht]
    \centering
    \includegraphics[width=\textwidth]{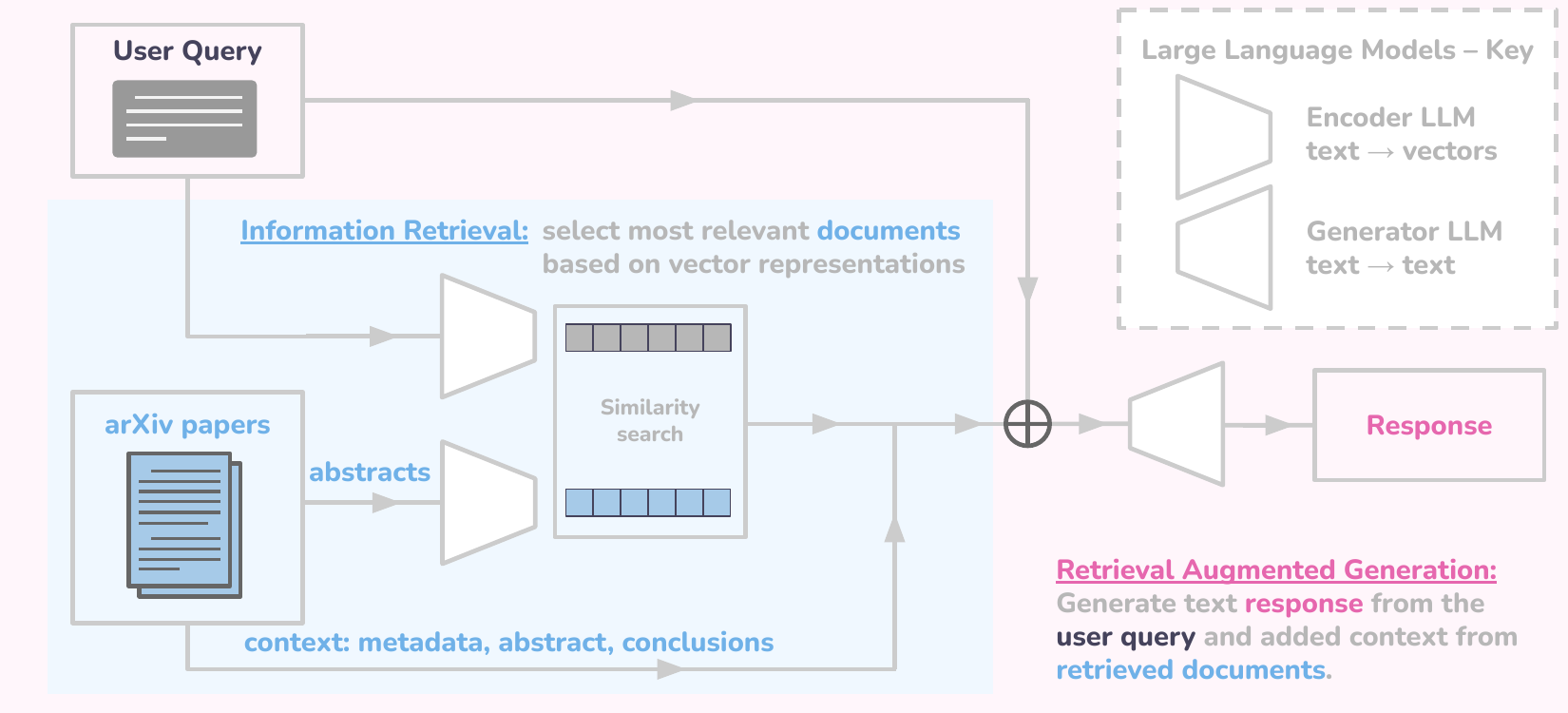}
    \caption{A schematic showing the LLM backend for our system. First, a user query is encoded and is used to retrieve $k=5$ similar papers based on their abstracts. After concatenating the prompt string, the top-$k$ papers' abstracts, conclusions, and metadata, and the original user query, we send it to the generator LLM, which outputs a response.}
    \label{fig:rag-schematic}
\end{figure*}

We build a RAG-powered LLM to respond to user queries based on the schematic in Figure~\ref{fig:rag-schematic}. First, the user query is encoded using the \texttt{bge-small-en-v1.5} encoder LLM \citep{bge_embedding}, and compared against a vector database of astronomy arXiv paper abstracts represented using the same encoder model. We use the arXiv astro-ph data set from \citet{Perkowski2024}, which comprises 300,000 arXiv papers up until July 2023 that were downloaded in \texttt{.tex} format and were subsequently cleaned. %

We select the top $k=5$ papers by cosine similarity, and combine the top papers' abstracts, conclusion sections, and metadata (arXiv IDs and years) into a context string. The context is concatenated with a prompt and the initial user query, allowing the LLM to send a reply using RAG. We currently use \texttt{gpt-4o} as the generator LLM. %

RAG can be sensitive to the wording of the user query, prompt, and retrieval hyperparameters like chunking or summarization. In our case, we do not perform any chunking or summarization of the arXiv paper abstracts because they always contain fewer than 1920 characters. Additionally, abstracts are generally in natural language (with limited markup), which improves the similarity search against the user's natural language query.

We find that curating the prompt leads to better RAG results. For example, the LLM often omits citations unless we include a strongly worded statement to always cite arXiv papers; we also provide a demonstration of this citation in the prompt. We prompt the LLM to prioritize more recent papers, making use of the paper's publication year in the retrieved context. Based on initial testing, these strategies appear to improve the LLM answers. 

\subsection{Slack chatbot interactions} \label{sec:slack-bot}

Our users will interact with the RAG-powered chatbot in the Space Telescope Science Institute (STScI) Slack workspace. Users can interact with the chatbot in two ways: by mentioning the chatbot (e.g., \texttt{@Ask astro-ph}) in a group channel where other users can also see messages, or via private direct messages (DMs) with the chatbot.
The Slack chatbot can only reply to user queries, and it interprets all messages as queries. Thus, we can treat the user interactions as simple question-answer pairs. Our server listens for user query messages via the \texttt{slack\_bolt} API.\footnote{\url{https://slack.dev/bolt-python/api-docs/slack_bolt/}} 

Users can upvote or downvote LLM answers by using emoji reactions, which triggers events that are recorded by the server. The Slack chatbot pre-populates two reactions, \texttt{:+1:} (thumbs up) and \texttt{:-1:} (thumbs down), which help guide the user toward upvoting or downvoting the LLM answer. We also allow users to leave any feedback they have regarding the model's response. Although our chatbot is not designed to handle multi-message interactions, additional data collected from users' feedback can serve as an insightful addition to the reaction data. %

\begin{figure}[t!]
    \centering
    \includegraphics[width=\columnwidth]{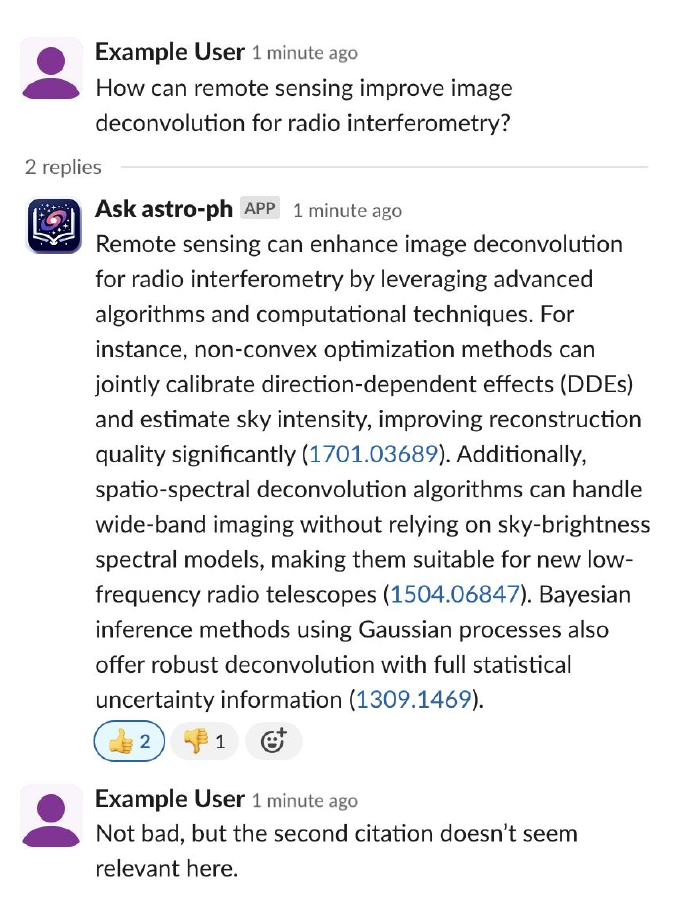}
    \caption{Example user interaction with the Slack chatbot.}
    \label{fig:slackbot}
\end{figure}

Figure~\ref{fig:slackbot} shows an example of interaction between ``Example User'' and the chatbot on Slack. In this case, the user sent a query via DM to the chatbot, and the chatbot replied with an answer in the message thread. This answer contains three citations to papers (with real hyperlinks). After the reply, the user gave a \texttt{:+1:} Slack reaction, indicating that the query was correctly answered, as well as a feedback message, which states that one of the citations appeared to be irrelevant to the query. Note that two \texttt{:+1:} votes are cast and one \texttt{:-1:} vote is cast because the Slack app pre-populates one of each reaction.

\subsection{Compiling and Annotating User Data}

\begin{figure*}
    \centering
    \includegraphics[width=0.75\textwidth]{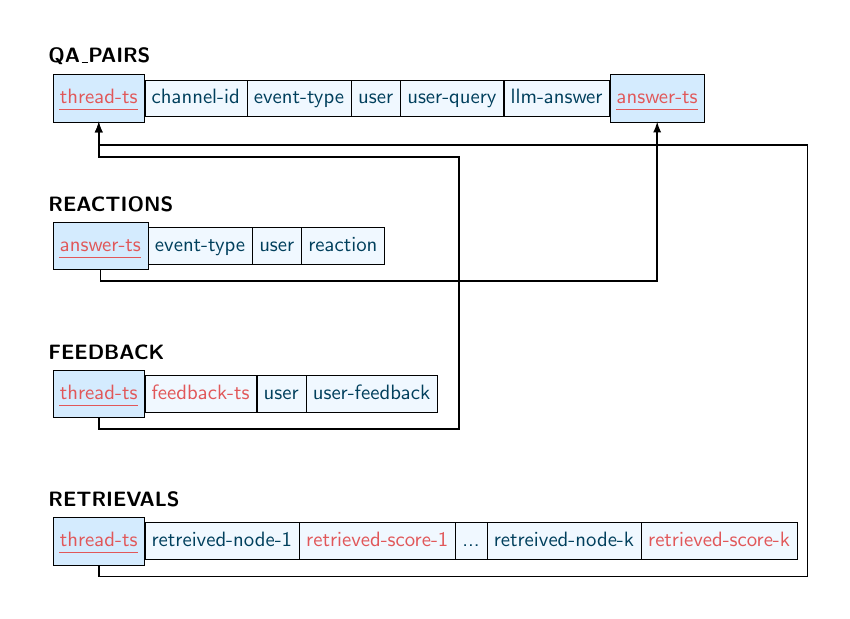}
    \caption{Table schema for user annotation data. Each box shows a column in a table, and arrows denote relationships between columns in different tables. Red and blue text color indicate data that are stored as floats and strings, respectively.}
    \label{fig:table-schema}
\end{figure*}

All data that will be collected are shown in the tables in Figure~\ref{fig:table-schema}. We note that Slack events are uniquely identified via timestamps. For example, user queries to the chatbot can be identified from its thread timestamp (\texttt{thread-ts}). 

When a user sends a query to the chatbot, a row will be written to the \texttt{QA\_PAIRS} and \texttt{RETRIEVALS} tables. The first table contains information about the channel and type of event, both of which will help determine whether the user sent a message via a Slack channel or DM. The user query and LLM answer are recorded here, as well as the unique answer timestamp.

The \texttt{FEEDBACK} table can have any number of rows per thread timestamp: any number of users can send any number of feedback messages. The \texttt{REACTIONS} table records a row every time a reaction is added or removed (under the column \texttt{event-type}) to the LLM answer. To count the number of upvotes, for example, we would filter by the specified reaction type (\texttt{:+1:}), and subtract the number of removed reactions from added reactions for all users.

\subsection{Optional demographic information}

After users sign the informed consent form to participate in our study, they are presented with an optional request for demographic information. We anticipate that our user base will be astronomy researchers with PhDs in physics or astronomy, but they may still have differing levels of seniority, research time availability, English proficiency, etc. This optional demographic information can help us develop a more holistic understanding of how astronomy users interact with LLMs.

\section{Towards LLM Evaluation for Astronomy} \label{sec:future-evaluations}

By using the aforementioned framework, we will be able to evaluate LLMs in a dynamic, real-world system (an active Slack workspace) that is heavily used by astronomy researchers. Moreover, our experiment will collect data that can be used to further improve future LLMs. However, we have not yet begun collecting data, so we can only present some preliminary topics of interest for later investigation. 

\subsection{Evaluating Research Topics}

Our data set will enable studies of how different user queries vary by research topic. Users might ask different types of questions depending on the astronomy subfield. For example, questions related to cosmology may request specific numerical values (e.g., ``What is the statistical significance of the Hubble tension?''), or perhaps questions related to exoplanets may feature named entities at higher rates than in other subfields (``Is there evidence that Kepler-22b is in the habitable zone?'').

The LLM answer quality may also vary with astronomy subfield. For example, RAG-based answers may struggle to form coherent responses to queries on hotly debated topics (e.g., ``Do major mergers trigger active galactic nuclei?''). LLMs may provide outdated or erroneous information more frequently for subfields with declining publication rates (i.e., such that the bulk of their papers are well in the past). A thorough study of the  \texttt{RETRIEVALS} table will be useful for characterizing LLM responses and failure modes.

We can also consider whether users ask different types of questions depending on the topic. For example, users may be more interested in seeking specific information for certain astronomy subfields, while requesting general background knowledge for other astronomy subfields. We note that such variations could be dependent on the particular user base that is being studied.

\subsection{User evaluation studies}

Our user data will also be essential for understanding astronomer preferences and LLM usage. For example, what fraction of users continue to ask questions after the first week of usage? How does usage change over time? Do (particular groups of) users primarily interact with the chatbot via private DMs, or do they mostly interact in a more public Slack channel? 

It will also be useful to study whether users find the LLM to be more useful over time (e.g., based on user reactions to answers). If this happens, then is it because only a select group of LLM-expert users are continuing to find it valuable? Or perhaps users are able to learn from each other, and thereby fashion queries that are more likely to give good answers?

We can also evaluate how user interactions depend on demographics, for users who opt in. We could test how an astronomer's seniority (years since PhD) or native language may correlate with LLM usage and successful interactions (e.g., as measured by a user who upvotes the LLM response to their original query).

\section{Conclusions}

We have presented a framework for dynamically evaluating how LLMs can be used in astronomy research (Section~\ref{sec:design}). We create a LLM-powered chatbot that cites information from astronomy arXiv papers, and we deploy the chatbot in a Slack workspace so that astronomers can interact with it. Through our experimental framework, we will record user questions and chatbot answers, user upvotes/downvotes to the LLM answer, open-ended user feedback, and retrieved papers and similarity scores. 

Although we have not yet begun collecting data, we introduce some prospective topics for detailed evaluation studies (Section~\ref{sec:future-evaluations}). These future evaluations can explore how user--LLM interactions depend on different astronomy subfields (e.g., exoplanets, interstellar medium, stars, galaxies, cosmology, or instrumentation). We also pose questions for evaluating how (or if) astronomers find LLMs to be useful.

Astronomy is an ideal proving ground for studying the potential benefits of LLMs to the scientific community, without danger of PII, societal risks, or commercialization. Evaluation studies will be crucial for understanding how astronomers interact with LLMs, and for improving future LLMs. This work introduces the evaluation framework for a study that will soon be under way. In a future paper, we will publish the evaluation data sets and our evaluation results.

\bibliography{main}
\bibliographystyle{icml2024}

\end{document}